\documentclass{article}
\usepackage{epsfig}
\usepackage{amsmath}
\usepackage{amsfonts}\tolerance=10000
\date{\today} 
 
\usepackage{graphicx}

\newcommand{\insertplot}[5]{\begin{figure}
 \hfill\hbox to 0.05in{\vbox to #5in{\vfill 
 \inputplot{#1}{#4}{#5}}\hfill}
 \hfill\vspace{-.1in}
 \caption{#2}\label{#3}
 \end{figure}}
 \newcommand{\inputplot}[3]{
 \special{ps: plotfile #1}
}
\newcounter{fig}

\newcommand{\beq}{\begin{equation}}
\newcommand{\eeq}{\end{equation}}
\newcommand{\beqs}{\begin{eqnarray}}
\newcommand{\eeqs}{\end{eqnarray}}

\numberwithin{equation}{section}
\newcommand{\be}{\begin{equation}}
\newcommand{\ee}{\end{equation}}
\newcommand{\bea}{\begin{eqnarray}}
\newcommand{\eea}{\end{eqnarray}}

\usepackage{graphicx}

\begin{document}

\title{Charged, rotating black holes in Einstein-Gauss-Bonnet gravity} 
 
 \author{
 {\large Yves Brihaye}$^{\dagger}$
\\ 
\\
$^{\dagger}${\small Physique-Math\'ematique, Universite de
Mons-Hainaut, Mons, Belgium}
}
 
\maketitle 
 
\begin{abstract}
We consider the Einstein-Gauss-Bonnet equations in five dimensions including
a negative cosmological constant and a Maxwell field. Using an appropriate Ansatz
for the metric and for the electromagnetic fields, 
we construct numerically black holes  with two equal angular momenta in the two orthogonal
space-like planes of space-time. Families of such solutions, labeled by the angular momentum
and by the electric charge
are obtained for many representative intervals of the Gauss-Bonnet coupling constant $\alpha$.  
It is argued that, for fixed values of $\alpha$, the solutions terminate into extremal black holes
at ($\alpha$-dependent) critical values of the angular momentum and/or of the electric charge.  
The influence of the Gauss-Bonnet coupling constant, of the charge and of the cosmological 
constant on the thermodynamics of the black holes and on their domain of existence is analyzed. 
\end{abstract}

\section{Introduction}
There has been  a huge activity in the study of fundamental 
interactions in space-times involving more than four dimensions in the last years. 
The standard Einstein-Hilbert action for $d$-dimensional space-times (with $d>4$)
is not the only choice leading to second order equations of motion.
Suitable combinations of terms with higher powers in the curvature can be added
such that second order field equations result. In the case $d=5$,
the most general theory of gravity possessing this feature is Einstein-Gauss-Bonnet (EGB)
gravity.  
Higher derivative curvature terms occur in many contexts such as in
semi-classical quantum gravity and in the effective low-energy
action of superstring theories. 
The  Gauss-Bonnet term appears as the first curvature stringy
correction to general relativity~\cite{1,Myers:1987yn} when assuming
that the tension of a string is  
large as compared to the energy scale of other variables.
Since black holes seem to be the most fundamental solutions of the conventional
four-dimensional theory of General Relativity,
the construction of such solutions in theories formulated in $d$-dimensions
and involving higher derivative curvature terms is a natural generalisation.
The inclusion of the Gauss-Bonnet term in the theory leads to new features of
the solutions (see e.g. \cite{Garraffo:2008hu,Charmousis:2008kc} for reviews and further
references). 
Due to the non-linearity of the equations, however, it is very difficult
to find non-trivial exact analytical solutions of the field equations 
with  higher derivative terms. The main solution existing in a closed form
is the Gauss-Bonnet counterpart of the Schwarzschild-Tangherlini black hole \cite{Boulware:1985wk,Wheeler:1985nh}.
Static Anti-de Sitter (AdS) black hole solutions in EGB gravity 
are also known in closed form presenting a number of interesting features 
(see e.g. \cite{rgcai,cno,neupane,torii} and references therein).
It is of interest to generalize these solutions to the rotating case.
In five dimensions, rotating solutions with a negative cosmological constant
have been considered in \cite{kim_cai} within a perturbative
approach. The numerical construction of rotating black holes (in $5$-dimensional EGB gravity) 
with equal angular momentum in two orthogonal
planes was addressed in \cite{bh_radu,bh_radu_ds} and \cite{bkkr}, respectively for
asymptotically Anti-de Sitter (AdS), de Sitter (dS) and flat space-time. In \cite{bh_radu,bh_radu_ds} the emphasis was put
on the construction of the action and the counterterms necessary to make the variational principle
and asymptotic charges well defined. 
Families of black holes were constructed and the physical parameters characterizing them
were estimated numerically. A systematic study of the domain of existence of the 
asymptotically flat black holes with respect
to their rotational parameter and the Gauss-Bonnet coupling constant is reported in \cite{bkkr}.
It turns out
that for fixed value of the Gauss-Bonnet coupling constant, the rotating solutions exist up
to a maximal value of the angular velocity at the horizon. The construction of the domain of existence
leads to numerical difficulties at the approach of the limiting extremal configuration.

In this paper, we extend the study of the domain of existence of the rotating black holes
in the presence of a negative cosmological constant (AdS case). 
For a vanishing cosmological constant, the  solutions exist for arbitrary
values of the Gauss-Bonnet coupling, which we call $\alpha$ in the following.
When a negative cosmological constant is supplemented to EGB gravity, the solutions exist only on
a limited  interval of $\alpha$. This occurrence of a maximal value for $\alpha$ is
examined within our approach of the field equations.
In addition to the purely gravitating EGB theory,
we further supplement the EGB action by a Maxwell term. Solving the corresponding
field equations we are able to construct families of
charged, rotating black holes of the Einstein-Gauss-Bonnet-Maxwell equations.
The influence of the electromagnetic charge on the domain of existence of the black holes is emphasized.

In section 2 we give the model, we present the Ansatz and also discuss the boundary
conditions. 
Several families of exact solutions are reviewed in section 3. Our results are
described in section 4 and illustrated by several figures. Section 5 gives a summary 
and conclusions.   

\section{The Model}
\subsection{The action and field equations} 
 We consider the Einstein-Gauss-Bonnet-Maxwell action with a cosmological constant $\Lambda$ given by
\begin{eqnarray}
\label{action}
I=\frac{1}{16 \pi G}\int_M~d^5x \sqrt{-g} \left(R-2 \Lambda+\frac{\alpha}{4}L_{GB} -  F^2  \right),
\end{eqnarray}
where $R$ is the Ricci scalar, $F^2$ is the standard Maxwell density and  $L_{GB}$ represents the Gauss-Bonnet density~:
\begin{eqnarray}
\label{LGB}
L_{GB}=R^2-4R_{\mu \nu}R^{\mu \nu}+R_{\mu \nu \sigma \tau}R^{\mu \nu \sigma \tau} \ .
\end{eqnarray}
In (\ref{action}), $G$ and  $\alpha$
denote the Newton constant and the Gauss-Bonnet coupling constant, respectively.  
It is usual to  relate the cosmological constant to the AdS radius $\ell$ 
according to $\Lambda = - (d-2)(d-1)/(2 \ell^2)$. 
The notations and conventions are the same as in \cite{bh_radu}.

The variation of the action (\ref{action}) with respect to the metric
tensor and the electromagnetic fields leads to the field equations 
\begin{eqnarray}
\label{eqs}
R_{\mu \nu } -\frac{1}{2}Rg_{\mu \nu}+\Lambda g_{\mu \nu }+\frac{\alpha}{4}H_{\mu \nu}= T_{\mu \nu} \ \ , 
\ \ \partial_{\mu}(\sqrt{-g} F^{\mu \nu}) = 0
\end{eqnarray}
where $T_{\mu \nu}$ is the electromagnetic stress-momentum tensor while the Gauss-Bonnet tensor is given by
\begin{equation}
\label{eq1}
H_{\mu \nu}=2(R_{\mu \sigma \kappa \tau }R_{\nu }^{\phantom{\nu}%
\sigma \kappa \tau }-2R_{\mu \rho \nu \sigma }R^{\rho \sigma }-2R_{\mu
\sigma }R_{\phantom{\sigma}\nu }^{\sigma }+RR_{\mu \nu })-\frac{1}{2}%
L_{GB}g_{\mu \nu }  ~.
\end{equation}
For a well-defined variational principle, one has to supplement the 
action (\ref{action}) with the Gibbons-Hawking surface term 
\begin{equation}
I_{b}^{(E)}=-\frac{1}{8\pi G}\int_{\delta \mathcal{M}}d^{4}x\sqrt{-\gamma }K~,
\label{Ib1}
\end{equation}
and its counterpart for Gauss-Bonnet gravity  \cite{Myers:1987yn} 
\begin{equation}
I_{b}^{(GB)}=-\frac{\alpha _{2}}{16\pi G}\int_{\delta \mathcal{M}}d^{4}x\sqrt{-\gamma }%
 \left( J-2\ {G}_{ab} K^{ab}\right)  ~,
\label{Ib2}
\end{equation}
where $\gamma _{\mu \nu }$ is the induced metric on the boundary,  $K$ is the
trace of the extrinsic curvature of the boundary,
  $\ {G}_{ab}$ is the Einstein tensor of the metric $\gamma _{ab}$ and $J$ is the
trace of the tensor
\begin{equation}
J_{ab}=\frac{1}{3}%
(2KK_{ac}K_{b}^{c}+K_{cd}K^{cd}K_{ab}-2K_{ac}K^{cd}K_{db}-K^{2}K_{ab})~.
\label{Jab}
\end{equation}
 
\subsection{The Ansatz}  
 While the most
general EGB rotating black holes would possess two independent
angular momenta and a more general topology
of the event horizon, 
we  restrict ourselves here to configurations with 
equal magnitude angular momenta and a spherical horizon topology.
The suitable metric Ansatz reads  \cite{Kunz:2005nm}
\begin{eqnarray}
\label{metric}
&&ds^2 = \frac{dr^2}{f(r)}
  + g(r) d\theta^2
+h(r)\sin^2\theta \left( d \varphi -w(r)dt \right)^2 
+h(r)\cos^2\theta \left( d \psi -w(r)dt \right)^2 
\\
\nonumber
&&{~~~~~~}+(g(r)-h(r))\sin^2\theta \cos^2\theta(d \varphi -d \psi)^2
-b(r) dt^2 \ \ ,
\end{eqnarray}
where $\theta  \in [0,\pi/2]$, $(\varphi,\psi) \in [0,2\pi]$, $r$ and $t$ being the
radial and time coordinates, respectively. 
For such solutions, the isometry group is enhanced from $R \times U(1)^{2}$
to $R \times U(2)$, where $R$ refers to the time translations.
This symmetry enhancement allows to transform the Einstein-Gauss-Bonnet
equations into a system of  ordinary differential
equations (ODEs). In our construction of the solutions, we fix the residual  freedom on the 
coordinate $r$ by taking $g(r)=r^2$. The  variable $r$ defined in this way is the counterpart
of the Schwarzschild radial coordinate. Note that in several papers dealing with black holes
in higher dimensions (see e.g.  \cite{Kunz:2005nm}) isotropic coordinates are adopted. 

When the Maxwell term is involved, the 
ansatz (\ref{metric})  has to be supplemented with a consistent parametrization of  
the electromagnetic field. The vector field is  parametrized
in terms of two functions according to 
\be
\label{amu}
  A_{\mu} dx ^{\mu} = V(r) dt + A(r) ( \sin^2 \theta d \varphi + \cos^2 \theta d \psi)
\ee
where $V(r)$, $A(r)$ represent respectively the electric  and  magnetic potentials. 
The ansatz (\ref{metric}), (\ref{amu}) transforms the set 
of Einstein-Gauss-Bonnet-Maxwell equations into
a system of six differential equations for the radial functions $f(r),b(r),h(r),w(r)$ and $V(r),A(r)$. 
These equations  can be combined in such a way that
the equation corresponding to the metric function $f(r)$ is of first order. 
The equation corresponding to the electric
potential depends only on $dV/dr$ and can be integrated in terms of the other functions.
For all the other functions the equations are of second order.
The form of these equations is involved and not given here; it can be found e.g. in \cite{bkkr,bh_del}.
In the following the prime will denote the derivative with respect to $r$. 

\subsection{Boundary conditions}
We  first discuss the regularity properties of the metric fields at the horizon
(the conditions  for the electromagnetic potentials will be given afterwards).
We are interested in solutions possessing a regular horizon at $r=r_h$. 
This requires the conditions $f(r_h)=0$, $b(r_h)=0$.
The fields can be expanded around the horizon leading to
\be
   f(r) = f_1 (r-r_h) + O(r-r_h)^2 \ \ , \ \ b(r) = b_1 (r-r_h) + O(r-r_h)^2 \ \ ,
\ee
\be 
   h(r) = h_0 + O(r-r_h) , \ \ w(r) = w_h + w_1 (r-r_h) + O(r-r_h)^2 \ ,
\ee
where $f_1,b_1, h_0$ are constants which are determined numerically while $w_h = w(r_h)$ 
(we also use  the notation $\Omega \equiv w_h$)
represents the angular velocity at the horizon. This parameter controls the rotation of the black hole and has to be
given by hand. The regularity of the fields at the horizon leads to an extra condition, say $\Gamma_1(f,b,b',h,h',w,w')_{r=r_h}=0$,  which needs to be imposed.
The polynomial $\Gamma_1$ is lengthy and is not given explicitly here.

In the asymptotic region, the metric is assumed to be 
asymptotically flat if $\Lambda = 0$ or Anti-de Sitter if $\Lambda < 0$.
The dominant terms of the asymptotic expansions  of the fields are (see \cite{bh_radu})
\be
\label{asymp}
f(r) =  \frac{r^2}{\alpha} \left( 1 - \sqrt{1- \frac{2\alpha}{\ell^2}} \right) + 1 + \frac{f_2}{r^2} + O(1/r^4) \ \ , \ \ 
b(r) =  \frac{r^2}{\alpha} \left( 1 - \sqrt{1- \frac{2\alpha}{\ell^2}} \right) + 1 + \frac{b_2}{r^2} + O(1/r^4) 
\ee
\be
h(r) = r^2 + \ell^2 \frac{f_2-h_2}{2r^2} \left( 1 + \sqrt{1- \frac{2\alpha}{\ell^2}} \right) + O(1/r^6)  \ \ , \ \ 
w(r) = \frac{w_4}{r^4} + O(1/r^8) 
\ee
Imposing the black hole condition, the regularity at the horizon, the asymptotic form of the metric and fixing
the angular velocity at the horizon through $\Omega$ leads to a consistent set of seven boundary conditions
specifying, in principle, a solution once $r_h$ and $\alpha$ are chosen.
The formula (\ref{asymp}) reveals that the expansion holds for $\alpha < \ell^2 /2$. In the limiting case
$\alpha = \ell^2/2$ the asymptotic expansions are different; e.g. 
the dominant term of the rotating
function $w(r)$ is $O(1/r^2)$
and for the metric function $h$ we have $h(r) = r^2 + H_0 + O(1/r^2)$. This contrasts with (\ref{asymp}).  
The corresponding model, called Chern-Simons gravity, was studied in \cite{MOTZ} and more recently
 in \cite{Anabalon:2009kq}.

In the presence of an electromagnetic field,
the boundary conditions for the functions $V(r)$ and $A(r)$ can be fixed as follows~:
at $r=r_h$ we set $V(r_h)=0$ and the condition  for the fields to be regular at the horizon leads to 
an equation
$\Gamma_2(f,b,b',h,h',w,w',V',A,A')_{r=r_h}=0$.
The polynomial $\Gamma_2$
(which comes out to be the same for the equations for $V(r)$ and $A(r)$
and is again too involved to be given explicitly)  
is independent from $\Gamma_1$ mentioned above.
In the asymptotic region, the electromagnetic potential has the following form 
\be
     V(r) = V_{inf} + \frac{q}{2 r^2} + O(1/r^4) \ \ , \ \ A(r) = \frac{q_m}{2 r^2} + O(1/r^4) \ ,
\ee
where $V_{inf}$, $q$ and $q_m$ are constants. In the absence of rotation (i.e. if $w(r)\equiv 0$) the field equation
corresponding to the magnetic potential  is trivially fulfilled by $A(r)=0$.
Practically,  we impose $V'(r\to \infty)= -q / r^3$ with a given value of the electric charge $q$.
The last boundary condition
consists in imposing  the product $r^2 A(r)$ to approach a constant.
The value of this constant, interpreted as the magnetic charge (say $q_m$) of the solution, depends on $\alpha,q,\Omega$ 
and can be extracted from the numerical data.  
Static solutions have $\Omega = q_m=0$.
\begin{figure}[h!]
\hbox to\linewidth{\hss%
	\resizebox{8cm}{7.1cm}{\includegraphics{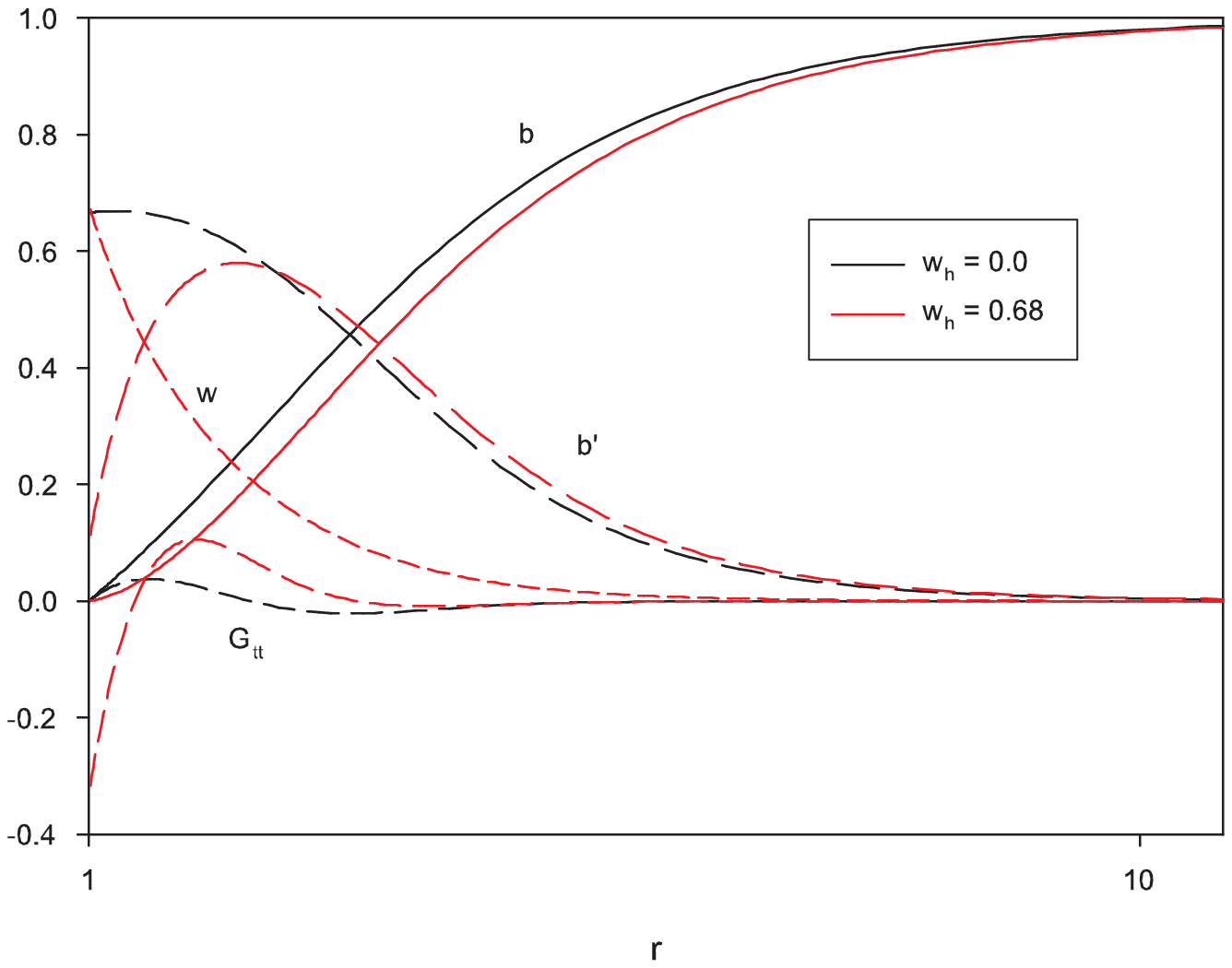}}
\hspace{5mm}%
        \resizebox{8cm}{7.1cm}{\includegraphics{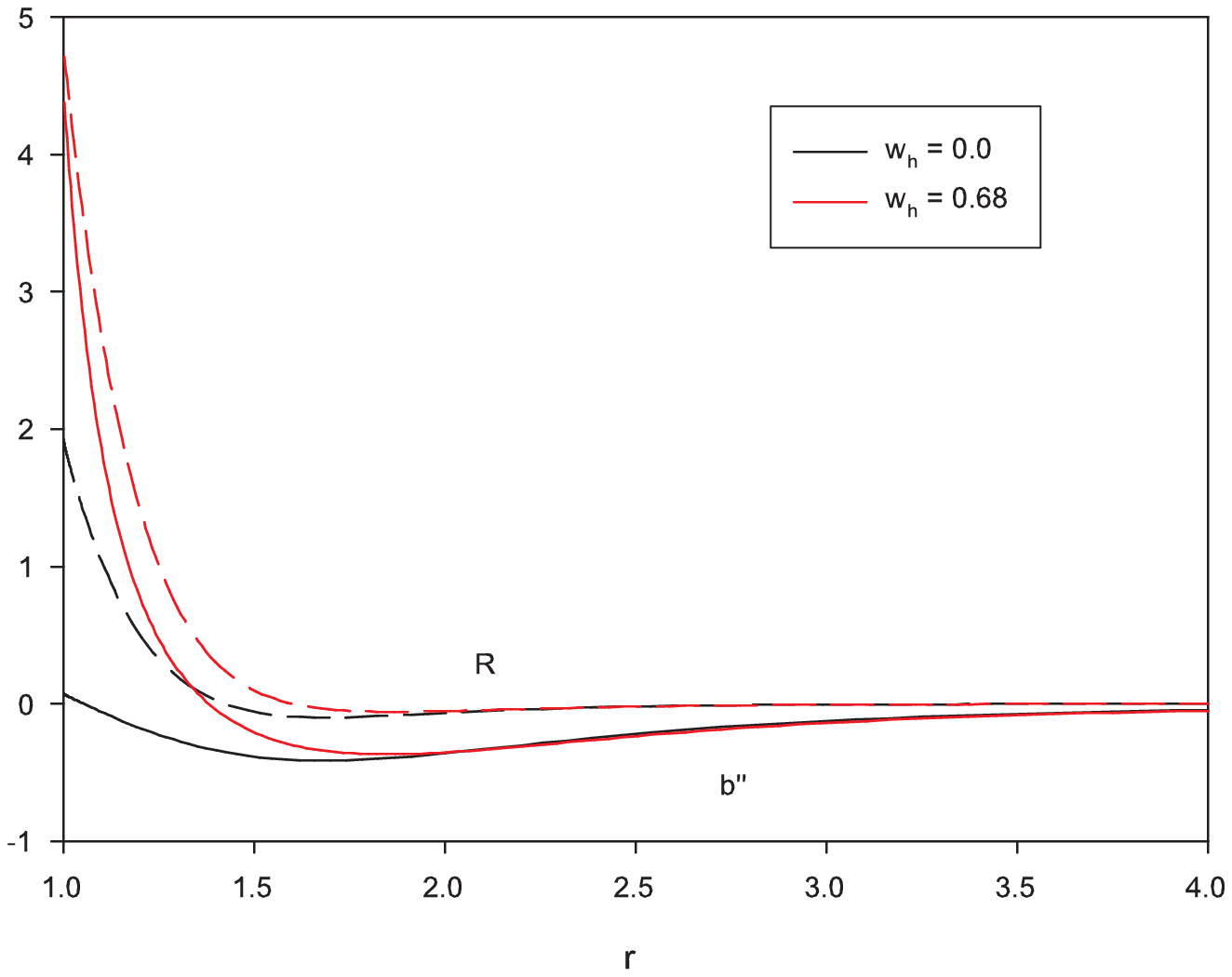}}	
\hss}
	\caption{  
Profiles of the solutions with $w_h=0.0$ and $w_h= 0.68$ for $\alpha = 2$ (left)	
and corresponding profiles of  $b''$ and of the Ricci scalar $R$ (right).
} 
\label{fig1a}
\end{figure}
\subsection{Physical quantities}
Physical quantities characterizing the black holes can then be extracted from the knowledge of the
different fields. 
In particular, the Hawking temperature $T_H$, the area $A_H$ of the horizon are given by
\be
             T_H = \frac{\sqrt{f_1 b_1}}{4 \pi}  \ \ , \ \ A_H = V_3 r_h^2 \sqrt{h(r_h)} \ . 
\ee  
The energy $E$ and the angular momentum $J$ are given by
\be
\label{energy_and_angular_momentum}
            E = \frac{V_3}{8 \pi G} \frac{f_2-4 b_2}{2} \sqrt{1 - \frac{2\alpha}{\ell^2}} \ \ , \ \ 
            J = \frac{V_3}{8 \pi G} w_4 \sqrt{1 - \frac{2\alpha}{\ell^2}} \ .
\ee 
In the figures in the next sections $T_H$ will be given in units of $4 \pi$ and $E,J$ 
in units of $V_3/(8 \pi G)$.
The electromagnetic properties of charged, rotating solutions can be characterized, namely by the electric charge $Q$, the
magnetic moment $\mu_m$ and the gyromagnetic ratio $g$
\be
\label{physical_em}
     Q = \frac{V_3}{8 \pi G} q \ \ , \ \ \mu_{m} = 
\frac{V_3}{8 \pi G} q_m \ \ , \ \ g = 2 \frac {E \mu_m}{Q J}  \ .
\ee 
Another relevant quantity for the analysis of the thermodynamics is the entropy $S$ of the black hole. 
It can be expressed in terms of the horizon $r_h$ and of the value $h_h \equiv h(r_h)$ of the metric function $h(r)$
at the horizon (see e.g. \cite{bh_radu}) according to
\be
             S = \frac{V_3}{4 G}\left(r_h^2 \sqrt{h(r_h)} + \alpha \sqrt{h(r_h)} 
\left(4 - \frac{h(r_h)}{r_h^2}\right)\right) \ \ .
\ee
The study of $S$ as a function of the temperature reveals some properties about the stability
through the specific heat $C_{\Omega}$ which is given by~:
\be
           C_{\Omega} = T_H (\partial S / \partial T_H)_{\Omega_H} \ .
\ee
\begin{figure}[h!]
\centering
\epsfysize=8cm
\mbox{\epsffile{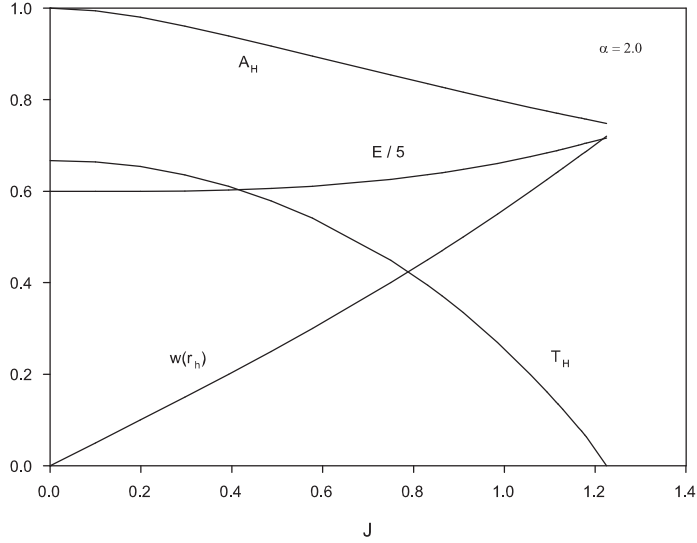}}
\caption{\label{fig3}
Some parameters characterizing the rotating solutions for $\alpha = 2.0$ as functions
of $J$.}
\end{figure}
 \section{Known solutions}
In this section, we review the explicit solutions of the equations which are available
in several specific limits. The fact that we recover these solutions with our numerical 
techniques provides a good crosscheck of the accuracy and robustness
of our method.
\subsection{Myers-Perry solutions}
We set $\Lambda = -6/\ell^2$ where $\ell$ is the AdS radius. 
For pure Einstein gravity, the Myers-Perry (MP) black holes \cite{Myers:1986un}  have the form
$$
   f = 1 + \frac{r^2}{\ell^2} - \frac{2M \Xi}{r^2} + \frac{2M a^2}{r^4} \ \ , \ \ 
h = r^2\left(1+\frac{2Ma^2}{r^4}\right) \ \ , \ \ b = \frac{r^2 f}{h} \ \ , \ \
   w = \frac{2Ma}{r^2 h}
$$
with $\Xi = 1  - a^2/ \ell^2$. The parameter $a$ is related to the angular momentum
 and $M$ is related to the horizon $r_h$ with $f(r_h)=0$.
Setting for simplicity $r_h=1$ 
the family of MP solutions present the following properties~:
$$
f'(r_h) = \frac{2(2a^2 (\ell^4 + 2 \ell^2 + 1) - \ell^4 -2 \ell^2 )}{\ell^2(a^2(\ell^2+1)-\ell^2)} \ \ , \ \ 
b'(r_h) = \frac{-2(2a^2 (\ell^4 + 2 \ell^2 + 1) - \ell^4 -2 \ell^2 )}{\ell^4} \ \ $$
$$ \ \ w(r_h) = a \frac{1 + \ell^2}{\ell^2} \ \ , \ \ M = \frac{1 + \ell^2}{2(\ell^2 - a^2(1+\ell^2))} \ .
$$
From now on we define $\Omega \equiv w_h \equiv w(r_h)$.
In particular, the solution is singular for $a^2=\ell^2/(1+\ell^2)$. 
The rotating BH are therefore limited to $a^2<\ell^2/(1+\ell^2)$.
Let  $\Delta \equiv a^2 - (\ell^2(\ell^2+2)/2(\ell^2+1)^2)$.
For $\Delta < 0$ the solutions have $f'(r_h) > 0$ and for $\Delta > 0$ they have $f'(r_h) < 0$
and admit a second horizon for $r = \tilde r_h > 1$. 
Since we want  to construct the solutions from the event horizon to infinity, 
it make sense to consider the
solutions only  for $\Delta < 0$ and for  $r \in [1, \infty]$. 
In the limit $\Delta = 0$ the black hole is extremal. 
The corresponding values of the parameters $a,M$, say $a_{ex}, M_{ex}$, are then fixed in terms of the AdS radius $\ell$~:
\be
a_{ex}^2 = \frac{\ell^2(\ell^2+2)}{2(\ell^2+1)^2} \ \ \ , \ \ \ M_{ex} = \frac{1+ \ell^2}{\ell^2} \ .
\ee 
The energy and angular momentum of the extremal black holes  can further be determined and read 
\be
     E_{ex} = \frac{V_3}{8 \pi G}\frac{6 \ell^4 + 13 \ell^2 + 8}{2 \ell^4}  \ \ \ ,\ \ \ 
     J_{ex} = \frac{V_3}{8 \pi G} \frac{\sqrt{2(\ell^2+2)}(1+\ell^2)}{\ell^3} \ \ .
\ee 
\begin{figure}[h!]
\hbox to\linewidth{\hss%
	\resizebox{8cm}{7.1cm}{\includegraphics{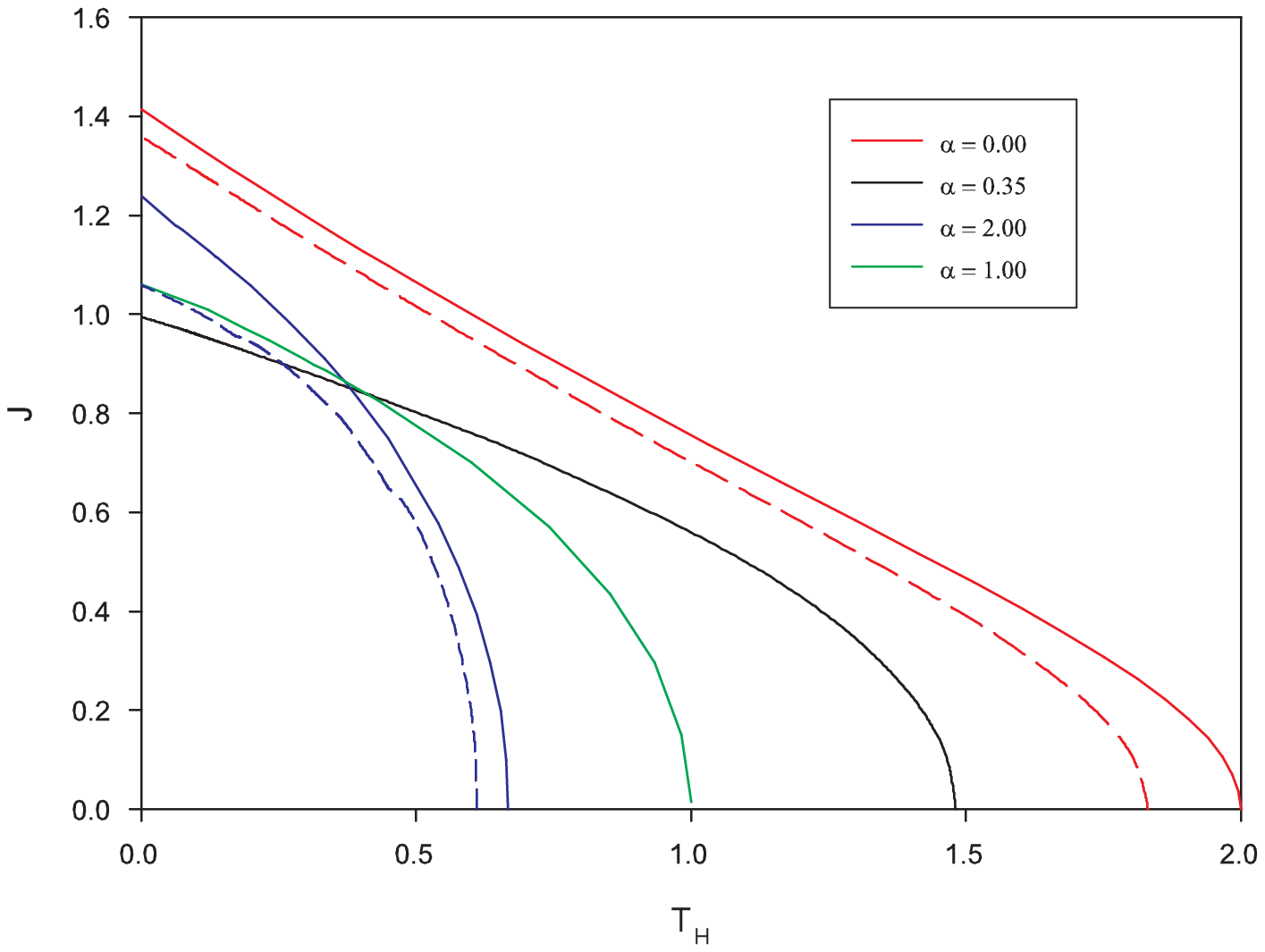}}
\hspace{5mm}%
        \resizebox{8cm}{7.1cm}{\includegraphics{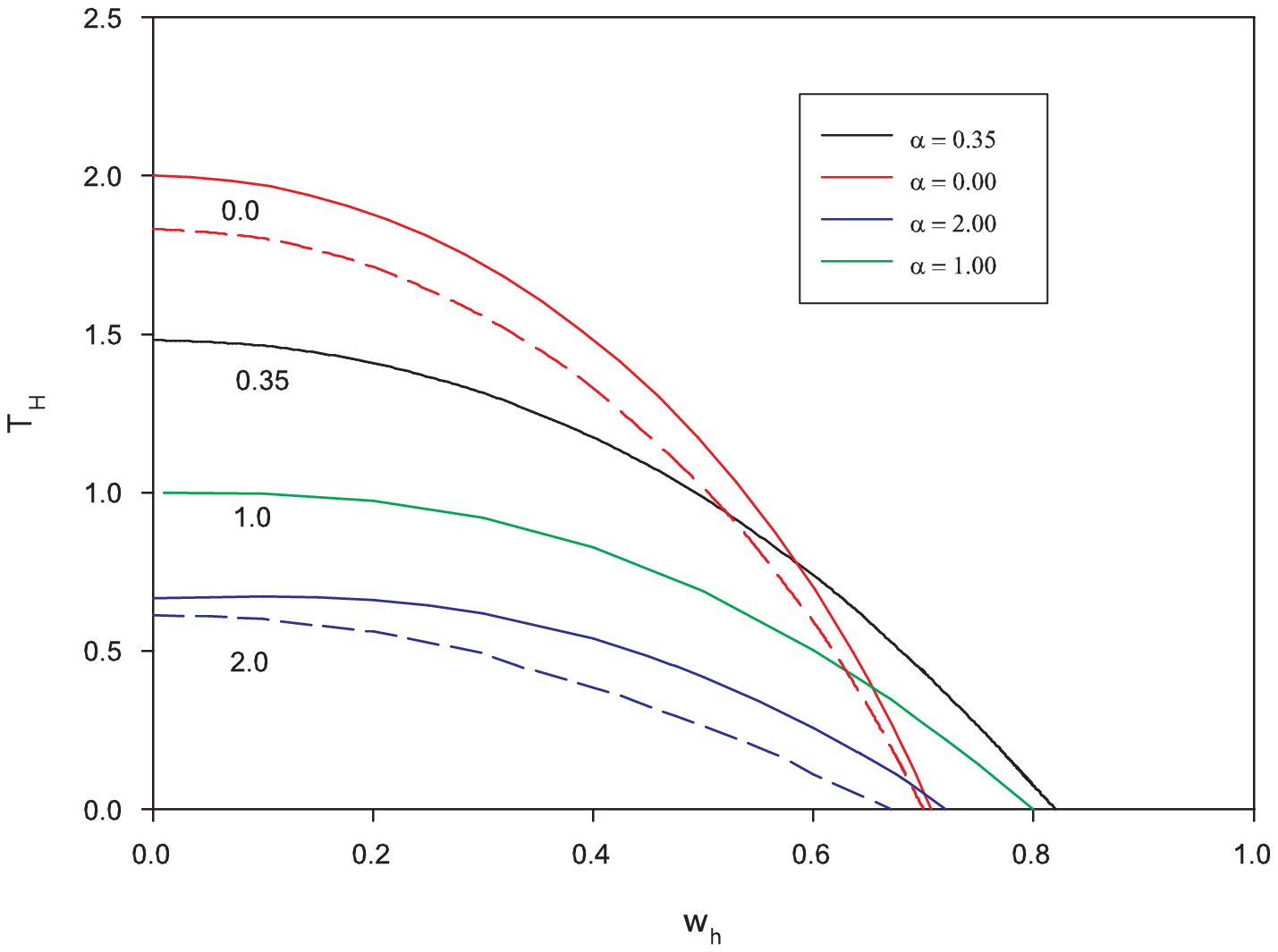}}	
\hss}
	\caption{  
Dependence of the angular momentum $J$ on the temperature $T_H$ for 
(asymptotically flat) rotating black hole with  
$\alpha=0,0.35,1,2$
(solid lines) and the charged counterparts (with $Q=1$, dashed lines) (left).
Dependence of $T_H$ on $w_h$ for rotating black hole with  $\alpha=0,0.35,1,2$
(solid lines) and the charged counterparts (with $Q=1$, dashed lines) (right).
} 
\label{J_TH_Q}
\end{figure}
\subsection{Anti-de Sitter-Reissner-Nordst\"om solutions}
Static charged black holes are known in an explicit form. 
In the limit $w(r)= A(r)=0$, the field equations
are solved by
\be
\label{rn}
f(r)=b(r) = 1 + \frac{r^2}{\ell^2} -\frac{2M}{r^{d-3}} + \frac{q^2}{2(d-2)(d-3)r^{2(d-3)}  } \
, \  h(r)= g(r)= r^2 \  , \  V(r)= \frac{q}{(d-3) r^{d-3}} \ ,
\ee
where $q$ is a parameter related to the electric charge.
Setting for definiteness $\ell^2=\infty$ and $r_h=1$, we find $M= (12+q^2)/24$ and $T_H = (2 - q^2/6)/4 \pi$.
The black hole is extremal when the charge become large enough: i.e. for $q = \sqrt{12}$.
\subsection{Static Einstein-Gauss-Bonnet solutions}
Concentrating now on the EGB equations and 
abandoning rotations (i.e. with $w(r)=A(r)=0$)  a family of exact solutions
\cite{Boulware:1985wk,Wheeler:1985nh,neupane,torii}
 exists with~: 
\be
\label{egb}
    f = b = 1 + \frac{r^2}{\alpha}(1 - \sqrt{1 + 2 \alpha (\frac{2 M}{r^4} - \frac{1}{\ell^2} - \frac{q^2}{12 r^6}) }) \ \ , 
    \ \ g = h = r^2 \ \ , \ \ w = 0 \ \ .
\ee
Setting $q=0$ and $\ell= \infty$ (i.e. $\Lambda = 0$) shows that
these solutions exist for arbitrary values of $\alpha$. If $\Lambda < 0$, 
they exist for  $\alpha \leq \frac{\ell^2}{2}$. 
Setting $r_h=1$, the charged EGB static black holes have
\be
      M = \frac{12+ 6 \alpha + q^2}{24} + \frac{1}{2 \ell^2} \ \ \ , \ \ \ 
      b'(r_h=1) = \frac{12- q^2}{6(1+\alpha)} + \frac{4}{ \ell^2(1+\alpha)} 
\ee
Knowing that black hole solutions exist in these two particular cases, $\alpha = 0$ and $\Omega = 0$
we  expect that  rotating black holes of the Einstein-Gauss-Bonnet equations should exist
in some domain of the $\alpha-\Omega$-plane. The purpose of the next section is to determine
this domain with some accuracy. 
\begin{figure}[h!]
\hbox to\linewidth{\hss%
	\resizebox{8cm}{7.1cm}{\includegraphics{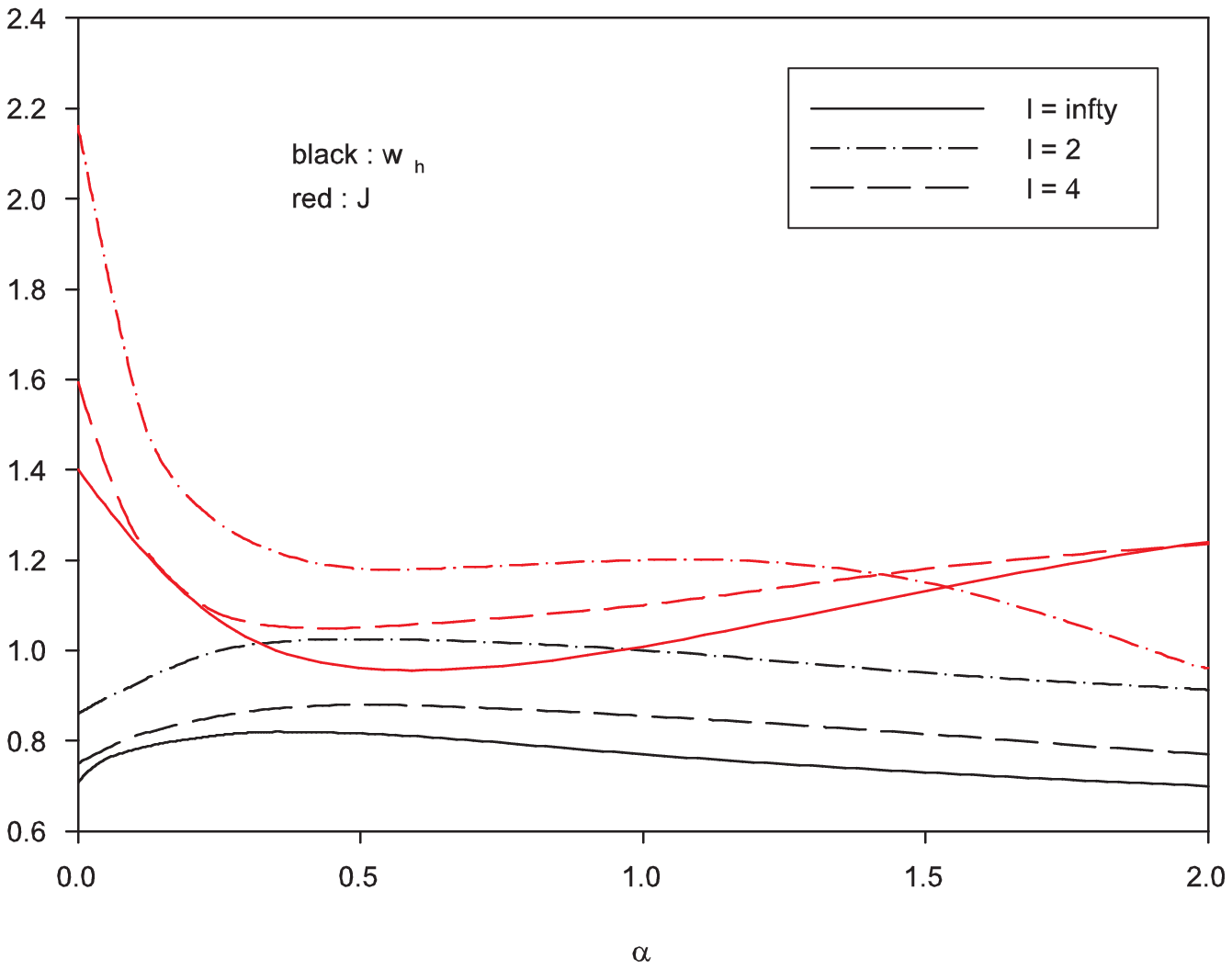}}
\hspace{5mm}%
        \resizebox{8cm}{7.1cm}{\includegraphics{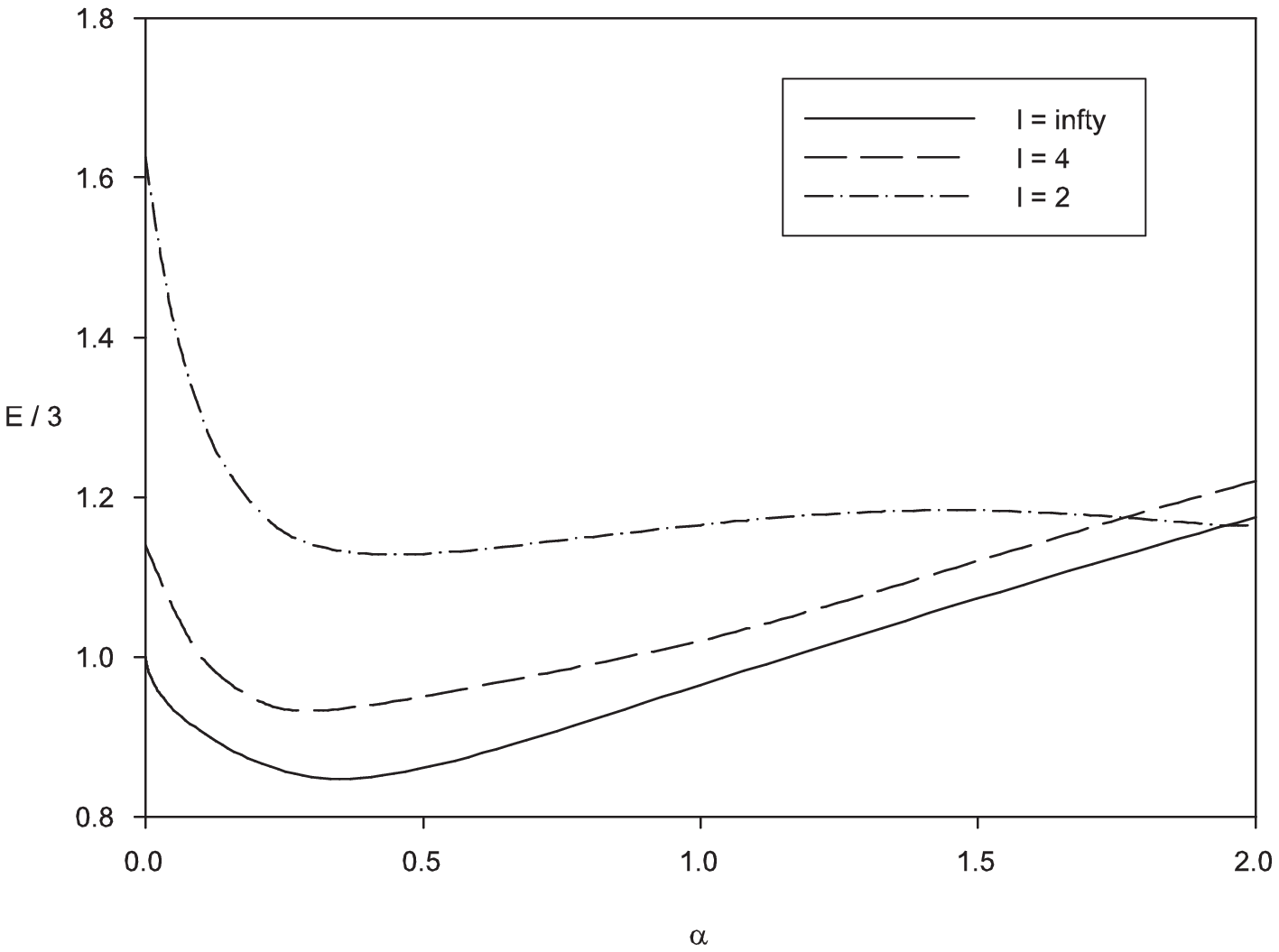}}	
\hss}
	\caption{  
Left :The parameters $w_m$ and  $J_m$ corresponding to the extremal black holes
as functions of $\alpha$ for several values of $\ell$. 
Right : The mass of the extremal black holes $M_m$ as function of $\alpha$ for several values of $\ell$
} 
\label{maximal_value}
\end{figure}
\section{Numerical results}
To our knowledge, no explicit solutions  exists in the generic case $w_h > 0$, $\alpha > 0$. 
The system of differential equations has to be solved numerically or perturbatively (e.g. using the
angular momentum as expansion parameter, see \cite{kim_cai}).
We solved the system numerically by using a collocation method \cite{colsys} for numerous
values of $\alpha$ and of $w_h$. 
For the main part of the analysis we set $r_h=1$ without loss of generality since this corresponds simply
to fixing the scale of the radial variable $r$. 
The study of the quantity 
        $b'(r_h,w_h) \equiv \frac{d b}{d r}|_{r=r_h,w(r_h)=w_h}$
(i.e. the value of the 
 derivative $db/dr$  for the solution with $w(r_h)=w_h$ at the horizon)
is of some interest for the understanding of the results.
Referring to the case $\alpha = 0$ and using an argument of continuity, 
it could be expected that 
a branch of rotating solutions exists for $\alpha >0$, stopping at a maximal value $w_{m}(\alpha)$
such that $b'(r_h, w_{m})=0$. This implies that an extremal black hole with $T_H= 0$ is approached. 
\begin{figure}[h!]
\centering
\epsfysize=8cm
\mbox{\epsffile{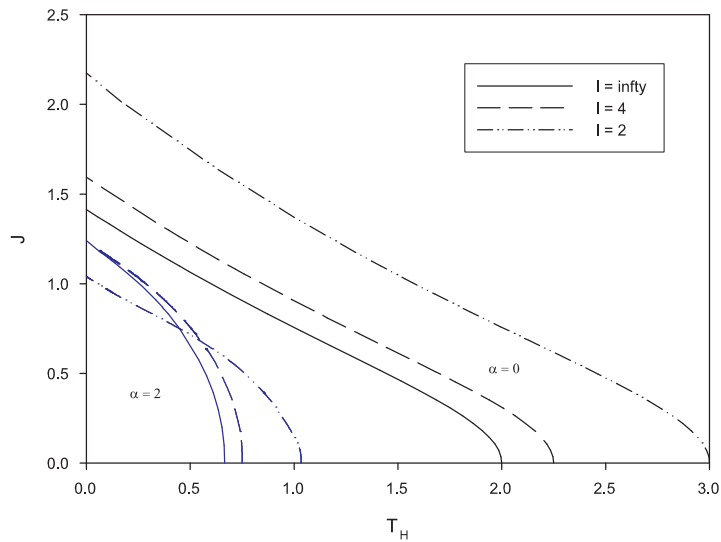}}
\caption{\label{fig4}
$T_H-J$ diagram for different values of $\alpha$ and of $\ell$.}
\end{figure}
\subsection{Asymptotically flat solutions.}
The solutions corresponding to $\Lambda = 0$ have been studied in \cite{bkkr},
but we find it useful to summarize the main features for completeness. 
Also, the deviation of the pattern of AdS solutions from the
asymptotically flat solutions will clearly appear. 
Fixing  $\alpha > 0$ and increasing $w_h$ gradually,
rotating black holes can be constructed up until a maximal value of $w_h$, i.e. $w_h = w_m(\alpha)$. 
Profiles of the solutions corresponding to $\alpha=2$ are presented in Fig. \ref{fig1a} for
the non-rotating solution $w_h=0$ and $w_h \approx 0.68$ 
(for $\alpha = 2$, we have $w_m \approx 0.72$).
\begin{figure}[h!]
\centering
\epsfysize=8cm
\mbox{\epsffile{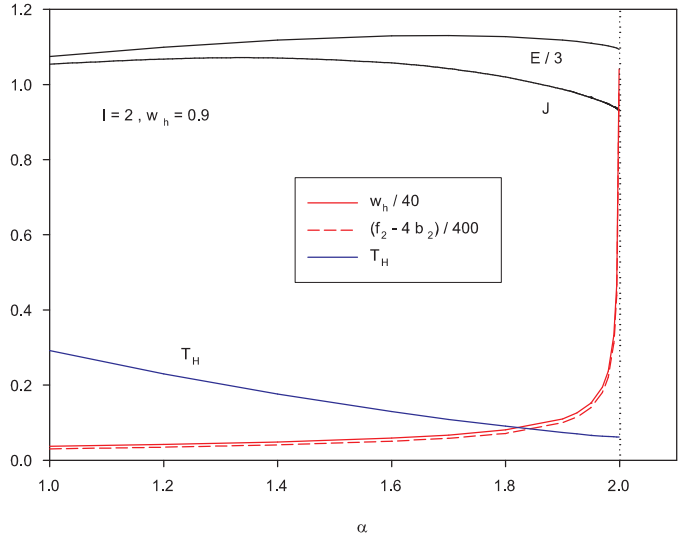}}
\caption{\label{chern_simons_limit}
Evolution of some parameters  with $\alpha$ for $\ell=2$ and $w_h=0.9$} 
\end{figure}
%
The study of the evolution of the parameters $b'(r_h)$, $f'(r_h)$ as  functions of $w_h$ suggests  a natural explanation 
for this result and reveals the pattern
of the solutions.  For $w_h=0$, we have  $b'(r_h=1)= 2/(1+\alpha)$, as can be computed 
from the explicit solution (\ref{egb}). Then, increasing $w_h$, our numerical results show that
$b'(r_h)$, $f'(r_h)$ both decrease monotonically and reach the value zero 
in the limit $w_h \to w_m$.
At the same time, the ratio $f''(r_h)/f'(r_h)$ becomes very large.
This critical phenomenon is illustrated in Fig.\ref{fig3} for $\alpha = 2.0$, where some   
physical quantities characterizing the solutions have been plotted as function
of the angular momentum. 
Reiterating this construction for different values of $\alpha$
confirms that the families of rotating solutions of the Einstein-Gauss-Bonnet equations 
systematically approach an extremal black hole for a maximal value of the 
angular velocity $w_h$. 
\begin{figure}[h!]
\hbox to\linewidth{\hss%
	\resizebox{8cm}{7.1cm}{\includegraphics{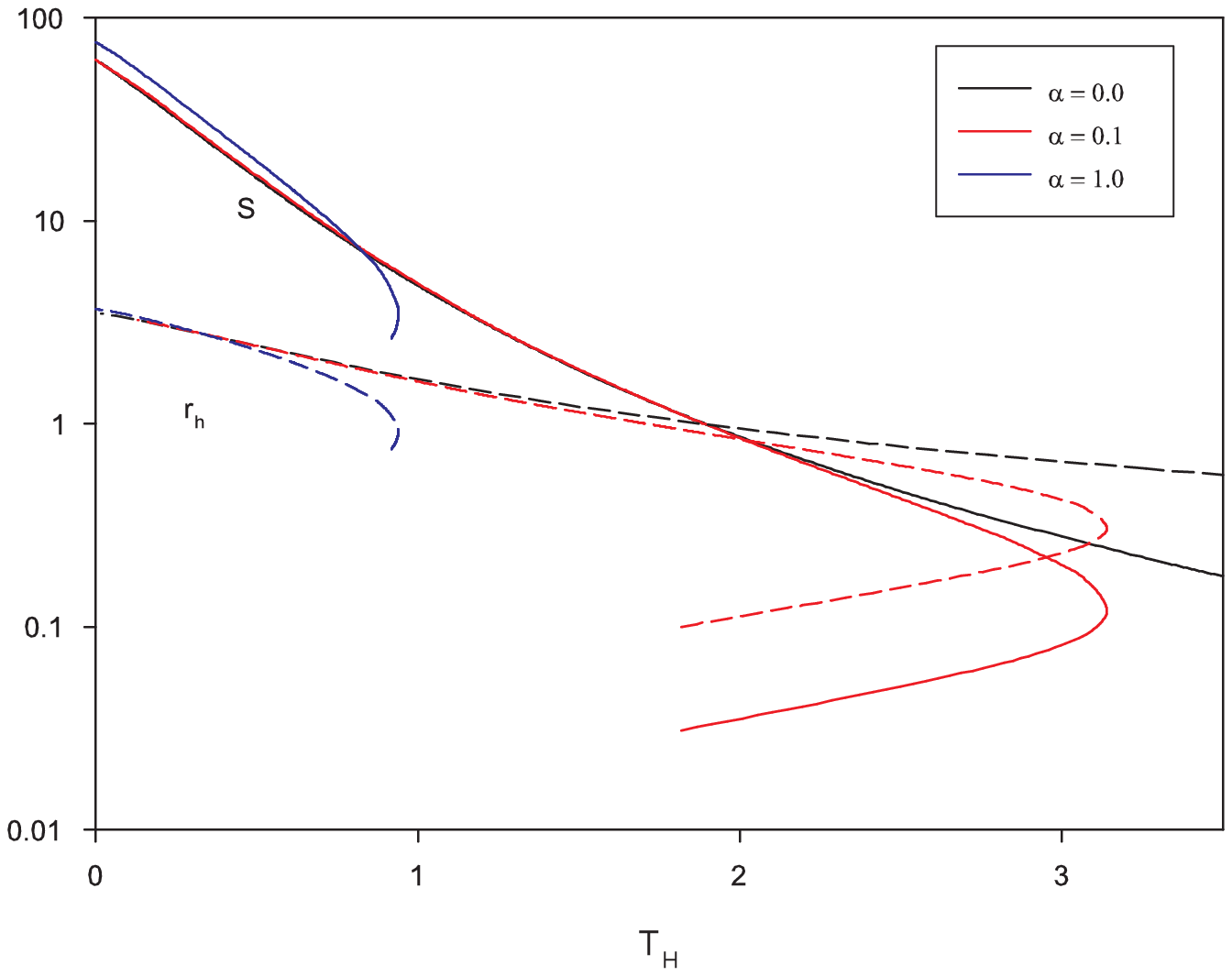}}
\hspace{5mm}%
        \resizebox{8cm}{7.1cm}{\includegraphics{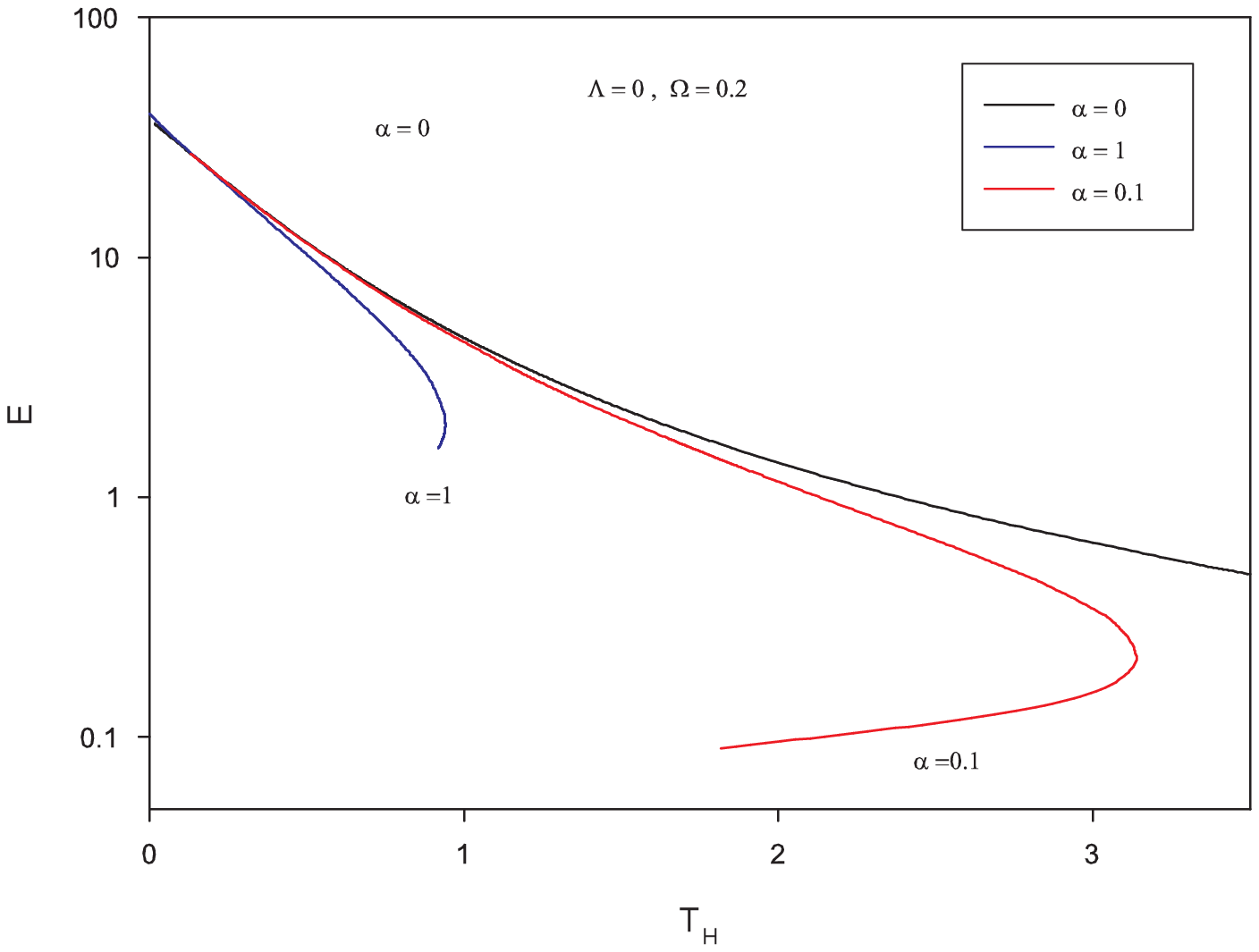}}	
\hss}
	\caption{  
The entropy $S$ and horizon $r_h$ as functions of the temperature 
for $\Lambda=0$, $\Omega=0.2$ and three
different values of $\alpha$ (left).
The mass as function of the temperature for $\Lambda=0$, $\Omega=0.2$ and three
different values of $\alpha$ (right).
} 
\label{thermo_1_entro}
\end{figure}
It is then natural to study how the physical properties of the black
holes are affected by  the Gauss-Bonnet parameter $\alpha$. 
This is illustrated in Fig. \ref{J_TH_Q}~: the left part  represents  
a temperature-angular momentum plot for four values of $\alpha$ (solid lines).   
The right part of  Fig. \ref{J_TH_Q} shows the temperature as a function of $w_h$.
Note that the lines $\alpha=0$ are determined analytically by means of
 $$
            T_H = \frac{2(1-2 w_h^2)}{\sqrt{1-w_h^2}} \ \ , \ \ 
J = \frac{w_h}{1-w_h^2}   \ \ , \ \ w_h \in [0, 1/\sqrt{2}] \ .
 $$  
The different curves show the non-trivial dependence on the solutions on $\alpha$. 
In particular $w_m$ and $J_m$
are not monotonic functions of $\alpha$. For the slowly rotating black holes 
the Hawking temperature decreases
while $\alpha$ increases. This statement clearly does not 
hold for the solutions with large angular momentum.
The natural question is then to attempt to  determine the domain of existence 
of rotating EGB black holes in the $\alpha-\Omega$-parameter space.
This is summarized in Fig. \ref{maximal_value} (left), where the 
values $w_m$ are plotted as a function of $\alpha$.
The angular momentum of the  extremal black hole is also given.
This domain turns out to be the region {\it below} the solid black line 
of the figure (representing $w_h$) 
or equivalently {\it below} the solid red line 
if we consider it from the point of view of the angular momentum $J$.
The corresponding energy of the limiting solution is plotted in Fig. \ref{maximal_value} (right).
The effect of the Gauss-Bonnet interaction is clearly perceptible in these plots.
The analysis was limited to $\alpha \in [0,2]$ since the Gauss-Bonnet 
term is supposed to come out
as a correction to the Einstein gravity. However, we found families of 
solutions presenting the same
pattern even for larger values of $\alpha$.
\begin{figure}[h!]
\hbox to\linewidth{\hss%
	\resizebox{8cm}{7.1cm}{\includegraphics{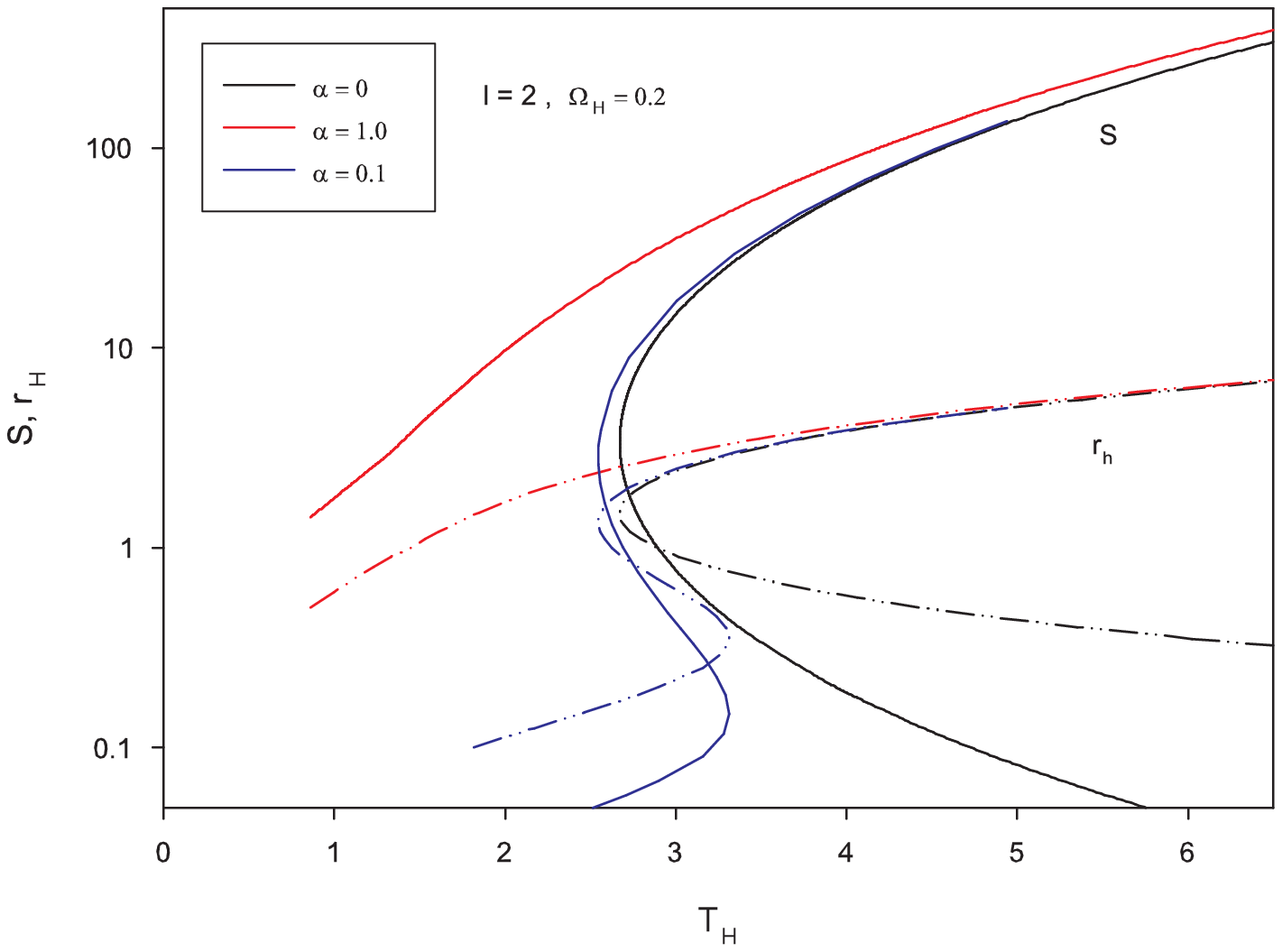}}
\hspace{5mm}%
        \resizebox{8cm}{7.1cm}{\includegraphics{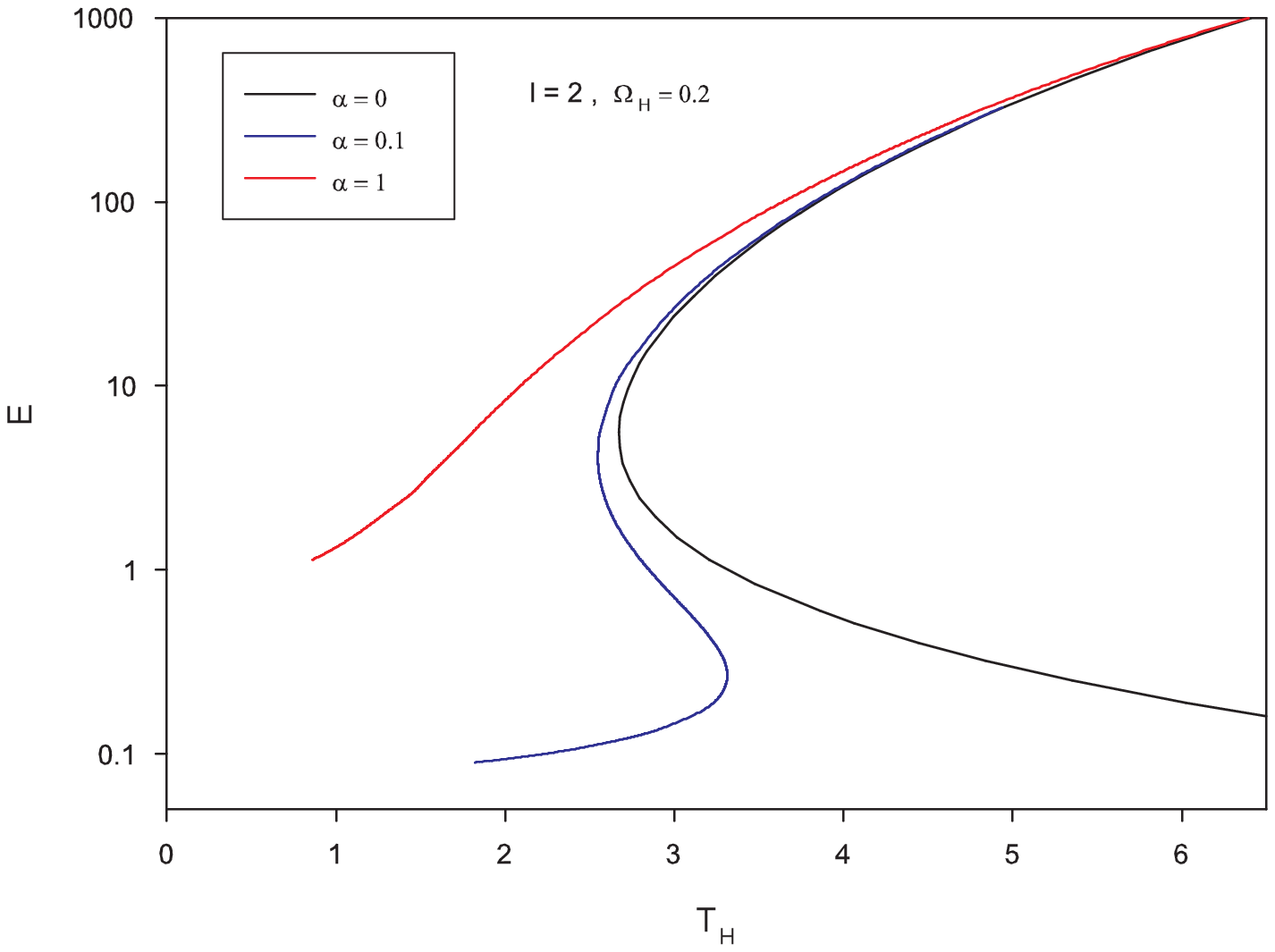}}	
\hss}
	\caption{  
The entropy $S$ and horizon $r_h$ as functions of the temperature for $\ell=2$, $\Omega=0.2$ 
and three
different values of $\alpha$ (left).
The mass as function of the temperature for $\ell=2$, $\Omega=0.2$ and three
different values of $\alpha$ (right).
} 
\label{thermo_2_entro}
\end{figure}
\subsection{Asymptotically Anti-De Sitter solutions}
We now discuss the asymptotically AdS  solutions (i.e. corresponding to  $\ell \ < \infty$).
The numerical results show that they approach continuously  
the asymptotically flat solutions in the limit $\ell \to \infty$. 
Fig. \ref{maximal_value}  contains the
data for $\ell = \infty$, $\ell=4$ and $\ell = 2$. 
The domain of existence of the solutions  varies smoothly with  the AdS radius $\ell$.
The influence of the cosmological constant on the temperature and angular momentum is
apparent in Fig. \ref{fig4}, where the curves are reported for two 
values of $\alpha$ and for $\ell = 2,4$ and $\ell = \infty$.
\begin{figure}[h!]
\hbox to\linewidth{\hss%
	\resizebox{8cm}{7.1cm}{\includegraphics{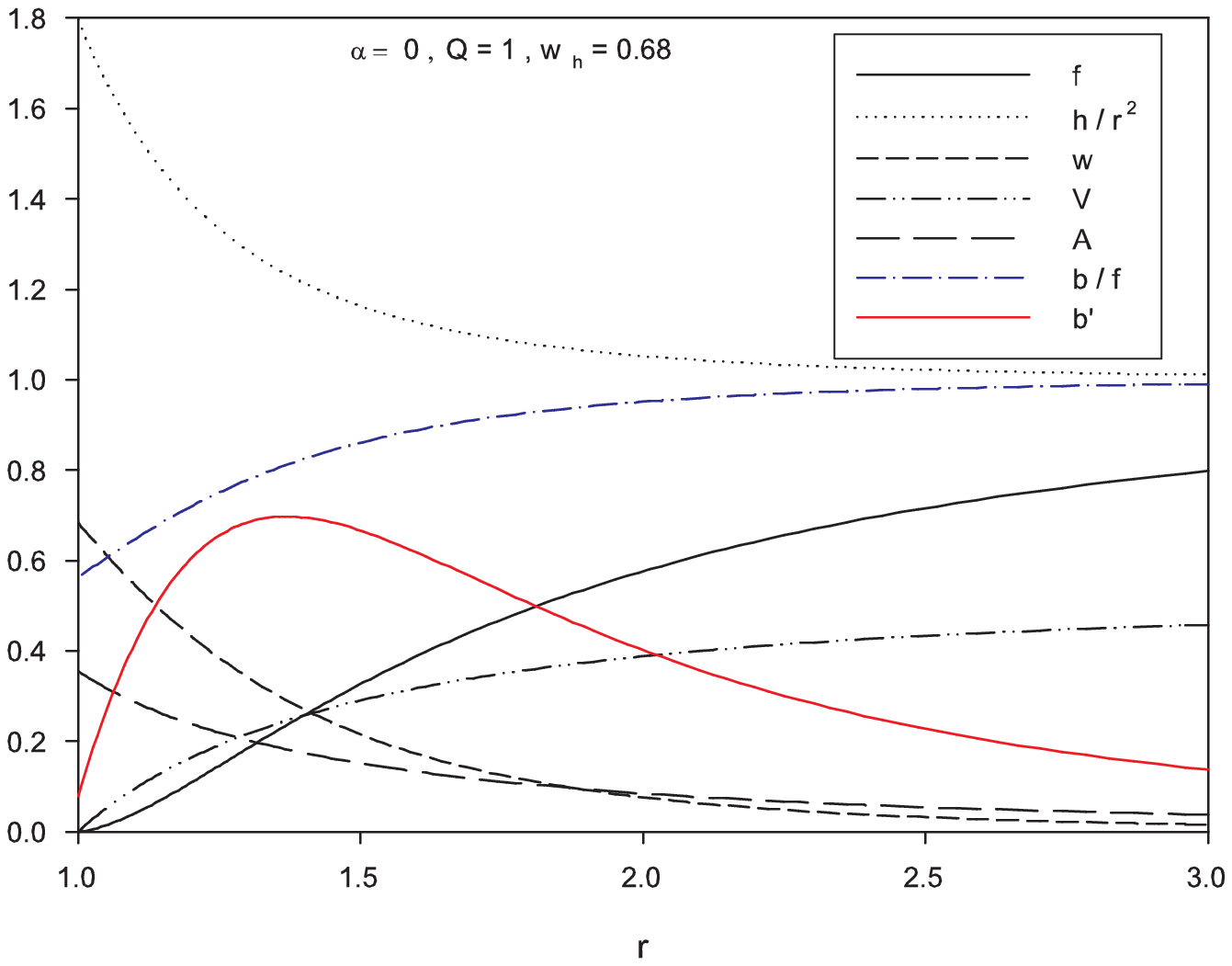}}
\hspace{5mm}%
        \resizebox{8cm}{7.1cm}{\includegraphics{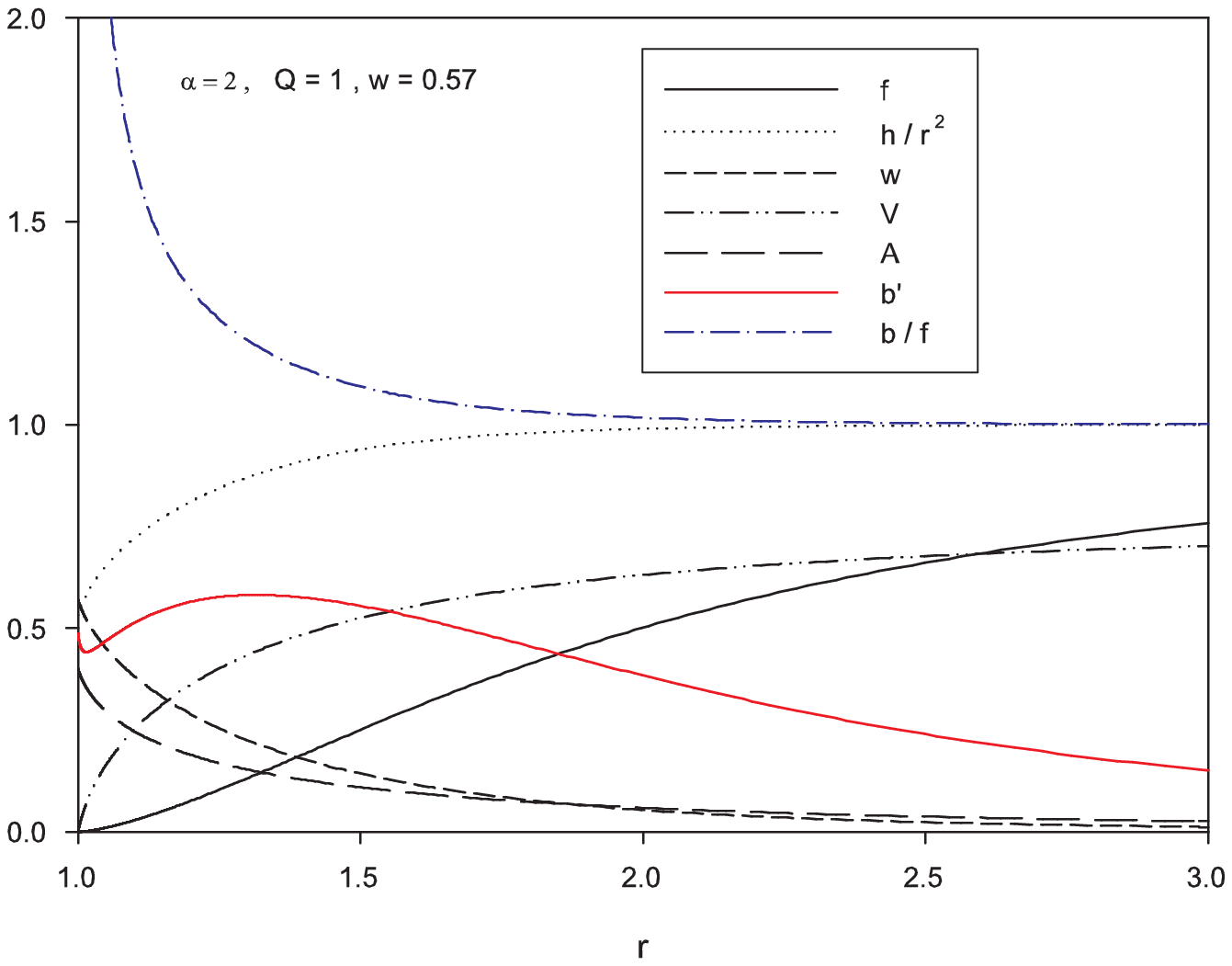}}	
\hss}
	\caption{  
Profile of a charged, rotating black hole with  $\alpha=0$ for $Q=1$ and $w_h=0.68$ (left).
and for  $\alpha=2$ for $Q=1$ and $w_h=0.68$ (right), respectively.
} 
\label{charged_bh_al_0}
\end{figure}
As pointed out already, the main difference between the flat and AdS black holes
resides in the fact that the asymptotically AdS
black holes exist only for $\alpha < \ell^2/2$, irrespectively of their angular momentum.
Constructing a branch of solutions with fixed $w_h$ while increasing $\alpha$ reveals that the 
asymptotic parameters $f_2,b_2$ and $w_4$ all diverge in the limit $\alpha \to \ell^2/2$
while the mass and angular momentum (\ref{energy_and_angular_momentum}) remain finite.
This is illustrated in Fig. \ref{chern_simons_limit}.
In fact the case $\alpha = \ell^2/2$ corresponds to Chern-Simons gravity \cite{MOTZ}. 
The solutions in this case will be discussed elsewhere \cite{BOR}.

\subsection{Thermodynamics}
Above we have discussed families of solutions with fixed event horizon $r_h$
and examined the influence of the other parameters on it.
In order to analyze the thermodynamics of the black holes, one usually 
fixes $\alpha$ and $\ell$
and constructs the families of solutions obtained by varying $r_h$. 
Here we emphasize the influence of the Gauss-Bonnet constant $\alpha$ and of 
the AdS radius $\ell$
on the pattern of the solutions. In this numerical construction, 
we keep the horizon angular velocity
$\Omega$ fixed (more precisely we set $\Omega = 0.2$). 
%
%
%
The case $\Lambda = 0$ was presented in detail in \cite{bkkr}. In
Fig. \ref{thermo_1_entro} we give the entropy (left) and the mass (right) of 
asymptotically flat black holes corresponding to three different values of $\alpha$.
As pointed out in \cite{bkkr}, the non-vanishing Gauss-Bonnet coupling constant
leads to a branch of stable black holes with $C_{\Omega} > 0$. These possess small
horizons. Note that it turns out  to be difficult to solve the equations for $r_h \to 0$.
We hence limit our analysis to $r_h > 0.1$. Our results confirm that no stable rotating black
holes exist for $\alpha = 0$ and that increasing $\alpha$ leads to the emergence of a stable
branch corresponding to small horizons. The branch of stable black hole gets larger
while $\alpha$ is increased.
These results contrast drastically with the corresponding asymptotically AdS black holes. 
The entropy and horizon radius are plotted in  Fig. \ref{thermo_2_entro}  for $\ell =2$.
The presence of a non-vanishing cosmological constant reveal a quite  different pattern.
For $\alpha = 0$ Fig. \ref{thermo_2_entro} reveals that black holes with a small horizon
are unstable. Increasing $r_h$, there is a phase transition and the solutions corresponding to
$r_h > 1.8$ are thermodynamically stable. 
Increasing gradually the parameter $\alpha$ reveals that the 
{\it unstable} branch corresponding to small black holes merges into a  
branch of {\it stable} solution.
Up to three solutions can be constructed which have the same angular velocity and same 
Hawking temperature
but different entropy and energy.
In particular we find two disjoint intervals of possible event horizon values with
$C_{\Omega} >0$. The first branch occurs for small horizon.
In this case we have  $C_{\Omega} < 0$ (i.e. unstable solutions)
for intermediate values of $r_h$ and the specific heat becomes again positive for large $r_h$.
Finally when $\alpha$ gets large enough (typically $\alpha = 1$), the unstable branch disappears 
and we find only one branch of stable solution. Interestingly, a similar 
thermodynamical pattern was observed in
\cite{KNLR} for charged, rotating solutions. Increasing the angular
momentum in that case leads to the different phases. 
By contrast, our solutions are not charged but the
different phases occur by changing the Gauss-Bonnet parameter.
\begin{figure}[h!]
\hbox to\linewidth{\hss%
	\resizebox{8cm}{7.1cm}{\includegraphics{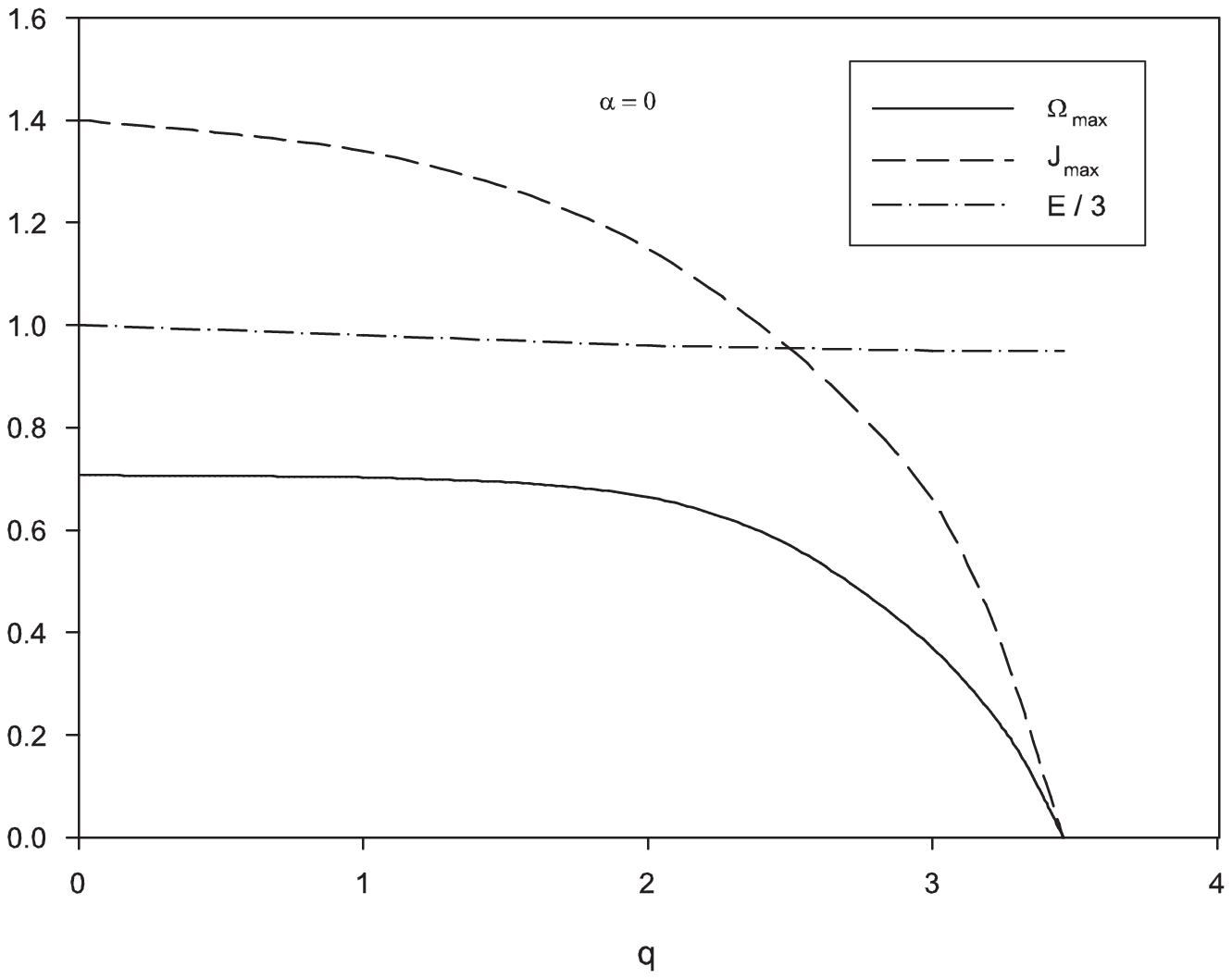}}
\hspace{5mm}%
        \resizebox{8cm}{7.1cm}{\includegraphics{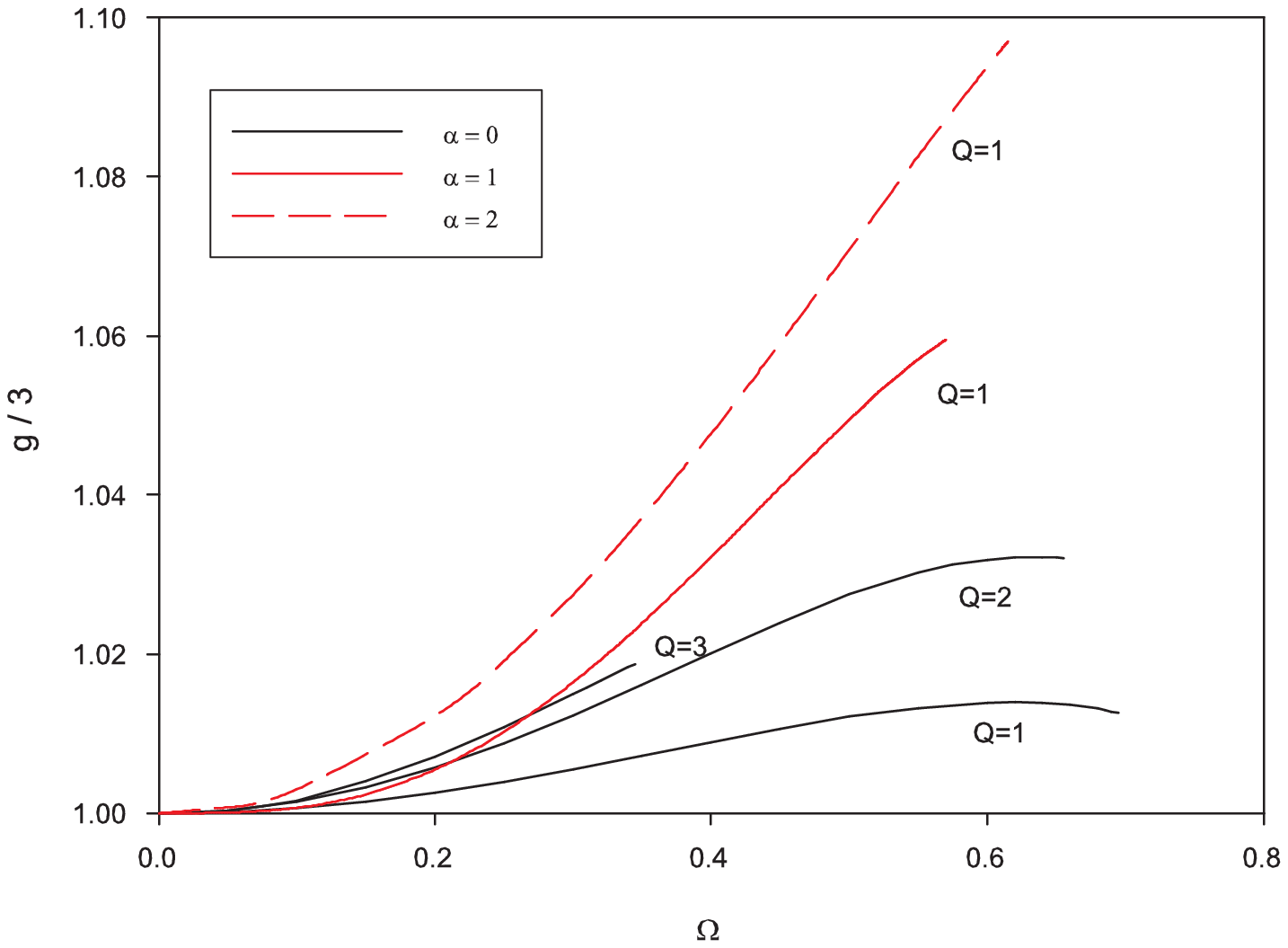}}	
\hss}
	\caption{  
Estimation of the values  $\Omega$, $J$ and $E$ corresponding to the extremal black hole 
as  functions of $q$ for $\alpha = 0$ (left).
Dependence of the gyromagnetic factor $g$ on $\Omega$ for several values of $Q$ and of $\alpha$
(right).
} 
\label{omega_q}
\end{figure}
The values of the energy of these families of black holes are given in Fig. \ref{thermo_2_entro} (right).
In the case when three solutions exist, it shows that the solutions with the smallest energy
correspond to the smallest horizon. 
%
%
\subsection{Charged solutions}
Because of the many parameters involved, we limit the construction of charged solutions
to asymptotically flat solutions (i.e. $\Lambda=0$) and fix the scale such that $r_h=1$.
Charged rotating black holes in Einstein-Gauss-Bonnet-Maxwell 
theory can then be constructed for different values
of the Gauss-Bonnet coupling constant $\alpha$, of the electric 
charge $q$ and of the angular velocity
at the horizon $\Omega$. 
The explicit solutions available in some special limits and the numerical results
suggest  these solutions to exist only in a limited domain of 
the $\alpha,q,\Omega$ parameter-space and to approach
extremal black holes on a surface
\be
           \Sigma \equiv S(\alpha,q,\Omega) = 0 \ .
\ee
The determination of this surface represents a formidable task but several 
limiting curves are available
from the special cases~: 
\begin{itemize}
\item  The set of points $(\alpha, \sqrt{12}, 0)$ belong to $\Sigma$.
\item  The critical line $(0,q,\omega_{max})$ is determined numerically 
and sketched in Fig. \ref{omega_q} (left).
Note that numerous solutions of the Einstein-Maxwell 
equations have been produced, e.g. in \cite{Kunz:2005nm},
using isotropic coordinates. The pattern of solutions might look different from ours since
we use a Schwarzschild-like radial coordinate and fix $r_h=1$.
\item  The critical line $(\alpha, 0, \omega_{max} )$ was obtained first in \cite{bkkr} 
and reproduced 
in Fig.\ref{maximal_value} given by the black-solid line on the left.
\end{itemize}
The surface $\Sigma$ can be expected to smoothly
extrapolate between these three limiting critical curves. 
Once the suitable boundary conditions are implemented, the equations
can be solved for generic values of $\alpha, \Omega, q$. 
For non-rotating black holes, imposing a charge by
means of $q \neq 0$ still leads to a zero magnetic potential $A(r)=0$. 
Charged-rotating black holes necessarily
have $A(r) \neq 0$ and are further characterized by a magnetic charge.
We produced indeed several families of charged, rotating black holes.
Fixing the electric charge $q$, our  results show that
the branch of rotating charged black holes exists and possesses
the same qualitative pattern as that of the uncharged solutions.
Profiles of charged-rotating black holes for $\alpha=0$ and $\alpha=2$  
are presented in Fig. \ref{charged_bh_al_0}.
Physical data characterizing families of rotating, charged solutions charge 
for several values of $\alpha$
are given in Fig. \ref{J_TH_Q}.
The solid lines corresponding to $q=0$ have been mentioned above.
The effect of the charge (we set $q=1$) is presented by the dashed lines for the cases 
$\alpha = 0$ and $\alpha = 2$ only. 
This figure reveals that the presence of the charge changes 
the domain of existence only little as compared to the uncharged black holes.
The numerical construction of several other branches 
(e.g. with $\alpha, \Omega$ fixed and varying $q$)
confirms this tendency and suggests the surface $\Sigma$ to be smooth. 
Finally, we studied some electromagnetic properties of the solutions. 
In particular, we present the
 response of the magnetic moment and gyromagnetic factor 
$g$ (see (\ref{physical_em})) on a change of the electric charge and of 
the Gauss-Bonnet coupling constant. This is summarized in Fig. \ref{omega_q} (right).
 It shows that the Gauss-Bonnet
term has a tendency to increase the factor $g$. As expected \cite{aliev}, 
the gyromagnetic factor tends
to $g = d-2$ (i.e. $g=3$) in the static limit. 
Note that the different curves on this figure stop at
the approach of the extremal black hole.
\section{Conclusions}
In this note we have presented arguments supporting the fact that equal angular momentum
rotating black holes of the Einstein-Gauss-Bonnet 
equations in five dimensions exist for generic values of the Gauss-Bonnet coupling constant.
In the case of a negative cosmological constant,
the asymptotically AdS solutions  exist up to a maximal value of the Gauss-Bonnet
coupling parameter, i.e. for $\alpha < \alpha_{cr} = \ell^2/3$. 
For $\alpha = \alpha_{cr}$ the model
corresponds to Chern-Simons gravity \cite{MOTZ}. The solutions obey a different asymptotic  
expansion than for generic values of $\alpha$. We were 
able to trace back this phenomenon with our
numerical construction and obtain robust solutions even for $\alpha = \alpha_{cr}$ 
by implementing
the appropriate boundary conditions.
\\ The effect of the Gauss-Bonnet term would be worth studying for other kinds of 
solutions of the underlying Einstein equations with a negative cosmological constant. 
It would be
challenging, in particular, to study the response of the 
NUT-charged solutions constructed in \cite{AMR}
to a Gauss-Bonnet term. Further applications of the Gauss-Bonnet black holes to the context of brane gravity
could be looked for by following, e.g. the ideas of \cite{bertha}.  
\\
Supplementing the EGB Lagrangian with a Maxwell term leads naturally to charged black holes.
The solutions terminate into 
extremal black hole at some maximal value of the angular momentum and/or of the electric charge.
We believe that the numerical estimates of the domain of existence of charged-rotating 
black holes in EGB gravity (which we studied  in the case $\Lambda = 0$ only) is 
correct qualitatively and quantitatively. This could be confirmed independently
by using an appropriate attractor technique along the lines of \cite{AGJST}.
\\
\\
{\bf\large Acknowledgments} 
This work is a continuation of a collaboration with Eugen Radu.
I gratefully  acknowledge him for many discussions. Thanks are also due to 
Kevin  Barb\'e for crosschecking some  numerical results and to B. Hartmann for reading the manuscript.
\cleardoublepage
 
\end{document}